\documentstyle[11pt,epsf]{article}
\begin{document}
\textwidth 15cm
\textheight 20cm
{\noindent{\large\bf Divergence Of Persistent Length Of A Semiflexible Homopolymer Chain In The Stiff Chain Limit}}

\vspace {.4cm}
{\noindent{\it\large Pramod Kumar Mishra}}
\vspace {.4cm}

{\noindent{\bf Department of Physics, DSB Campus, Kumaun University, Naini Tal-263 002 (Uttarakhand) India.}}
\vspace {.2cm}

{\noindent Email: pkmishrabhu@gmail.com}
\vspace {.2cm}

\noindent {\bf Abstract} :

In this brief report, we revisit analytical calculation [Mishra, {\it et al.}, Physica A {\bf 323} (2003) 453 
and Mishra, NewYork Sci. J. {\bf{3(1)}} (2010) 32.]
of the persistent length of a semiflexible homopolymer chain in 
the extremely stiff chain limit, $k\to0$ (where, $k$ is stiffness of the chain) for directed walk lattice model
in two and three dimensions. Our study for two dimensional (square and rectangular) 
and three dimensional (cubic) lattice case clearly indicates that the persistent
length diverges according to expression $(1-g_c)^{-1}$, where $g_c$ is the critical value
of step fugacity required for polymerization of an infinitely long linear semiflexible 
homopolymer chain and nature of the 
divergence is independent of the space dimension.
This is obviously true because in the case of extremely stiff chain limit
the polymer chain is a one dimensional object and its shape is like a rigid rod.

\vspace {.2cm}


\vspace {.2cm}


\vspace {.2cm}
\section{Introduction}
The persistent length of a polymer chain measures correlations 
in the orientation of the segments of the chain along its length.
In other words, the persistent length
is a measure of a distance along the chain length at which the configuration of the chain on an 
average has memory of the orientation of its specific segment. 
The bending rigidity and thus the persistent length is a consequence 
of short range atomic and molecular interactions present in the polymer chain. Since,
the persistent length is stemming from the bending rigidity of the polymer chain 
and it can exhibit enormous variation in the magnitude. Therefore, if 
persistent length associated with the polymer chain is much smaller 
than the overall length of the chain, such a chain is said to be flexible and
stiffness of such chain is unity. 
When stiffness of the chain is approaching to zero, the persistent length
of such chain being comparable to it's length and the chain is said to be rigid. 
However, if stiffness of the chain has value in between 0-1, the chain is said to be semi-flexible.
Actin filaments, microtubules, 
$DNA$, $protein$ and collagen are the examples of the semiflexible polymers. 
The persistent length plays
an important role in describing elastic properties of a semiflexible polymer chain 
and also plays vital role in developing theory of polyelectrolytes 
solutions. 

Due to excluded volume effect, self avoiding polymer chain has memory of it's specific segment  
and initial bias persists along the walk of the chain upto a finite distance (for flexible chains) from initial step of the chain. Grassberger \cite{1}
initially discussed this problem and showed that the persistent length of a two dimensional self avoiding 
flexible polymer chain
diverges with power law. Later, Redner and Privman \cite{2} suggested that this divergence 
is logarithmic. However, through MC studies \cite{3} it has been
shown that the persistent length could be fitted by power law and by a logarithmic function.
Eisenberg and Baram \cite{4} demonstrated and confirmed that the persistent length of a flexible polymer
chain converges to a finite value. The situation is different in the case when polymer chain 
is semiflexible and in the extremely stiff chain limit the persistent length of a semiflexible 
polymer chain diverges.

The aim of present report is to take into account correlations prevailing between two distant segments
of an extremely rigid polymer chain of an infinitely long length in the bulk and to demonstrate through simple calculations 
that the persistent length of
such polymer chain when expressed in terms of critical value of step fugacity in the extremely stiff chain limit (i. e. $k\to0$) diverges as a simple pole and the nature of divergence 
is independent of space dimensionality. 

This report is organized as follows: In Sec. 2,
we define directed walk model in brief and revisit the results of calculation of
the persistent length for
two dimensional (square and rectangular) and three dimensional 
(cubic) lattice to investigate the divergence of the persistent length 
of an infinitely long linear semiflexible homopolymer chain in the extremely stiff chain limit. 
Finally, in Sec. 3, we conclude the discussion by summarizing the results obtained.

\section{Model and method of calculations}

We consider following two  
cases of directedness \cite{5} of the polymer chain for square, rectangular and cubic lattices:
In the case (i) partially directed self avoiding walk ($PDSAW$) model,
the walker is allowed to walk along $\pm y$ and $+x$ directions on a square or a rectangular lattice while
in the cubic lattice case walker is allowed to walk along $\pm y$, $+x$ and $+z$ 
directions. In case (ii) fully directed self avoiding walk ($FDSAW$) 
model, the walker is allowed to take steps along $+x$, $+y$ directions in the square and rectangular lattice
case while along $+x$, $+y$ and $+z$ directions for the case of a cubic lattice. A partially directed self avoiding walk
is shown graphically on a two dimensional rectangular and a square lattice in figure (1).
\begin{figure}[htbp]
\epsfxsize=13cm
\centerline{\epsfbox{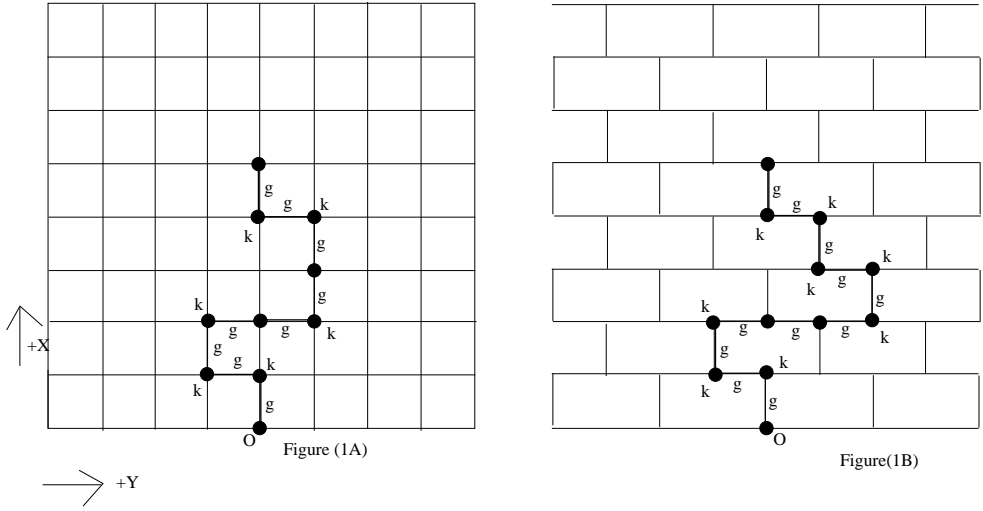}}
\caption{In this figure a partially directed self avoiding walk of a linear semiflexible polymer chain is shown (1A) on a
square lattice of 9 steps and (1B) on a two dimensional rectangular lattice of 11 steps. The step fugacity of each step is shown by $g$, 
$k[=exp(-\beta\epsilon_b)]$ is the stiffness of the polymer chain, $\epsilon_b$ is the value of bending energy required to produce
one bend in the chain and $\beta[=\frac{1}{k_BT}]$ is inverse of thermal energy. 
The Boltzmann weight of the walk shown in figure (1A) is $g^9k^6$ and of (1B) is $g^{11}k^8$. }  
\label{Figure5}
\end{figure}


The partition function of the chain is defined as follows: 
\begin{equation}
Z(g,k)=\sum_{N=0}^{N=\infty}{\hspace{.4cm}}\sum_{all\hspace{.2cm}walks\hspace{.2cm}of\hspace{.2cm}N\hspace{.2cm}steps}g^Nk^{N_b}
\end{equation}
Where, $N_b$ is the number of bends in a walk of a polymer chain of $N$ steps (monomers) and $g$ is fugacity associated with each
step (monomer).
The partition function of the chain is calculated \cite{6,7} by us using method of generating function technique \cite{5}.

The persistent length is defined by Mishra {\it et al.} \cite{6},
as an average length of the polymer chain between two successive bends, ${\it i. e.}$
$l_p=<L>/<N_b>=(g\frac{\partial Log[Z(g,k)]}{\partial g})/(k\frac{\partial Log[Z(g,k)]}{\partial k})$, where length of the chain is  
$L(=Na$, $a$ being the lattice parameter and $N$ is number of monomers in the chain). We have taken value
of lattice parameter unity for mathematical sake. 


\subsection{$PDSAW$ model on a square lattice:}

The partition function of a linear semiflexible homopolymer chain for this model is written as 
$Z_{PD-S}(g,k)=\frac{(4k-3)g^2+3g}{1-2g+g^2-2g^2k^2}$, \cite{6}, where $g$ is step fugacity and $k$ is stiffness 
weight associated with each bend of the polymer chain.

The critical value of step fugacity required for polymerization of an infinitely long linear semiflexible homopolymer
chain is determined from the singularity of the partition function. The critical value of step fugacity for 
partially directed self avoiding walk model of 
the chain on a square lattice is written in terms of $k$ as, 
$g_c= \frac{1}{1+\sqrt 2 k}$ $\cite{6}$. This allows us to write $k$ in terms of $g_c$ as, $k=\frac{1-g_c}{\sqrt{2}g_c}$. 

The persistent length of the polymer chain for $PDSAW$ model on a square lattice can be written as $\cite{6}$,
\begin{equation}
l_p= \frac{3+2\sqrt{2}}{4+3\sqrt{2}}[\sqrt{2}+\frac{1}{k}]
\end{equation}

Substituting $k=\frac{1-g_c}{\sqrt{2}g_c}$ in Eq. (2), we obtain expression of the persistent length as,
 
\begin{equation}
l_p= {(1-g_c)}^{-1}
\end{equation}

\subsection{$FDSAW$ model on a square lattice:}

For fully directed self avoiding walk model on a square lattice the partition
function of the chain is written as 
$Z_{FD-S}(g,k)=\frac{2g}{1-(1+k)g}$, \cite{6} while $g_c=\frac{1}{1+k}$, $\cite{6}$. Therefore,
we have expression for $k$ in terms of $g_c$ as, $k=\frac{1-g_c}{g_c}$, while
persistent length  for
this case is, $l_p= 1+k^{-1}$, $\cite{6}$. Substituting the value
of $k$ in terms of $g_c$ for this case too, we get,
\begin{equation}
l_p= (1-g_c)^{-1}
\end{equation}

\subsection{$PDSAW$ model on a two dimensional rectangular lattice:}
We have considered a rectangular lattice which has lattice parameter one unit along $x-$axis and
two unit along $y-$axis. This rectangular lattice can be derived from a two dimensional hexagonal lattice and the
lattice is shown in figure (1B).  
The partition function of the polymer chain  for this case is written as \cite{7}:

$Z_{PD-R}(g,k)=\frac{3g+2g^2+2g^2k-g^3+4g^3k-4g^3k^2}{1-g^2-2g^2k^2}$

In the case of a two dimensional rectangular lattice, the critical value of step fugacity
for polymerization of an infinitely long linear semiflexible homopolymer chain is written 
in terms of $k$ as, $g_c=\frac{1}{\sqrt{1+2k^2}}$, $\cite{7}$. In other words, $k$ in terms of $g_c$ is
written as $k=\frac{1-{g_c}^2}{2{g_c}^2}$, while the persistent length 
has dependence on $k$ as, $l_p=1+\frac{1}{2k^2}$
for $PDSAW$ model on a rectangular lattice. The persistent length (on substitution of $k$ in terms of $g_c$)
is re-written in terms of $g_c$ as,

\begin{equation}
l_p= (1+g_c)^{-1}(1-g_c)^{-1}
\end{equation}

\subsection{$FDSAW$ model on a two dimensional rectangular lattice:}
The partition function of the polymer chain for this case is 
$Z_{FD-R}(g,k)=\frac{2g+g^2+g^2k-g^3+2g^3k-g^3k^2}{1-g^2-g^2k^2}$, \cite{7} and we have,
$g_c=\frac{1}{\sqrt{1+k^2}}$, $\cite{7}$ from the singularity of the partition function. 
In this case, $k$ in terms of $g_c$ is
written as, $k=\frac{1-{g_c}^2}{{g_c}^2}$ and $l_p=1+\frac{1}{k^2}$ for $FDSAW$ model on a 
rectangular lattice in two dimensions. 
On substitution of $k$ in terms of $g_c$ for $FDSAW$ model on a two dimensional rectangular lattice,
we get,

\begin{equation}
l_p=(1+g_c)^{-1}(1-g_c)^{-1}
\end{equation}

\subsection{$PDSAW$ model on a cubic lattice:}
Thr partition function of the polymer chain for partially directed self avoiding walk model is 
$Z_{PD-C}(g,k)=\frac{(6k-4)g^2+4g}{(1+k-4k^2)g^2-(k+2)g+1}$, \cite{6}.
In this case the persistent length of the polymer chain is written as $\cite{6}$, 

\begin{equation}
l_p=\frac{2u_1[k^{-2}+k^{-1}-4]}{(1-\sqrt{17}+2k^{-1})
u_2+(85+21\sqrt{17})k^{-2}}
\end{equation} 

where

$u_1=85+19\sqrt{17}-(102+26\sqrt{17})k^{-1}+(34+8\sqrt{17})
k^{-2}$

and $u_2=204+52\sqrt{17}-(272+64\sqrt{17})k^{-1}$.

The critical value of step fugacity for this case is $g_c= \frac{k+2-\sqrt{17}k}{2(k+1-4k^2)}$, $\cite{6}$. For this case too, we follow the method
discussed above and substitute, $k=\frac{(1-g_c)(\sqrt{17}-1)}{8g_c}$ to obtain,

\begin{equation}
l_p=(1-g_c)^{-1}
\end{equation}
In this case  dependence of the persistent length on $k$ (as shown in Eq. (7)) is more involved than 
the cases discussed in sub-sections (2.1-2.4) and expression of the persistent length reduces to a simple form, 
as we have discussed in sub-sections (2.1-2.4),
when the persistent length is expressed in terms of $g_c$ 
{\it i. e.} Eq. (8). 
\subsection{$FDSAW$ model on a cubic lattice:}
The partition function of the polymer chain for $FDSAW$ model on a cubic lattice
is written as 
$Z_{FD-C}(g,k)=\frac{3g}{1-(1+2k)g}$, \cite{6}. 
The critical value of step fugacity is $g_c= \frac{1}{(1+2k)}$ and the persistent length is
$l_p=1+\frac{1}{2k}$ $\cite{6}$ for $FDSAW$ model on a cubic lattice. In this
case too (on substitution of $k=\frac{1-g_c}{2g_c}$ in the expression of the persistent length) we obtain,
\begin{equation}
l_p=(1-g_c)^{-1}
\end{equation}

\section{Conclusions}

We have used definition of Mishra {\it et al.} $\cite{6}$ to investigate
nature of the divergence of the persistent length
of an infinitely long linear semiflexible homopolymer chain in the extremely stiff chain limit {\it i. e. } $k\to 0$.
In this limit the polymer chain
is a one dimensional object and average length of the polymer chain between its two successive bends diverges
as, $(1-g_c)^{-1}$. 
In other words, the persistent length diverges as $l_p\sim(1-g_c)^{-1}\sim\frac{1}{k^q}$ (where, $q$ is an integer) for extremely stiff chain limit. 

When persistent length is expressed in terms of $k$, the constant of proportionality will depend on lattice dimension
and model. The constant of proportionality will have different value for isotropic model to that of 
directed walk model. However, when persistent length is expressed in terms of $g_c$,
we expect that the nature of the divergence of an average distance between two successive bends of the polymer chain  
will remain same for directed and undirected self avoiding walk models and constant of proportionality will have
different value for isotropic (undirected) model than the directed walk model. 
The nature of divergence is identical for partially and fully
directed walk models of the polymer chain for two and three dimensional lattices.
This is due to fact that in the extremely stiff chain limit the polymer
chain is a one dimensional object and its shape is like a rigid rod.

The qualitative nature
of variation of the persistent length with stiffness of the chain has similar variation for directed and
isotropic self avoiding walk models in two and three dimensions. However, exact value of the persistent
length of the chain will depend on space dimensions and type of model (directed or isotropic) chosen to enumerate walks of the chain \cite{8}.  


\small
\begin{thebibliography}{99}
\bibitem{1} P. Grassberger, Phys. Lett. A {\bf 89} (1982) 381.
\bibitem{2} S. Redner and V. Privman J. Phys. A: Math. Gen. {\bf 20} (1987) L857.
\bibitem{3} H. Meirovitch, J. Chem. Phys. {\bf 79} (1983) 502; 
H. A. Lim and H. Meirovitch, Phys. Rev. A {\bf 39} (1989) 4176;
D. E. Burnette and H. A. Lim, J. Phys. A: Math. Gen. {\bf 22} (1989) 3059.
\bibitem{4} E. Eisenberg and A. Baram, J. Phys. A: Math. Gen. {\bf 36} (2003) L121.
\bibitem{5} V. Privman and N. M. Svrakic, {\it Directed models of polymers, interfaces and clusters: Scaling and finite size properties}
(Springer, Berlin), (1989). 
\bibitem{6} P. K. Mishra, S. Kumar and Y. Singh, Physica A {\bf 323} (2003) 453.
\bibitem{7} P. K. Mishra, New York Sci. J. {\bf 3(1)} (2010) 32.
\bibitem{8} D. Giri, P. K. Mishra and S. Kumar, Ind. J. Phys. A {\bf 77(1)} (2003) 561.
\end{thebibliography}
\end{document}